\documentclass[aps,pre,twocolumn,noshowpacs,noshowkeys]{revtex4-2}

\usepackage{graphicx}
\usepackage{dcolumn}
\usepackage{bm}
\usepackage{hyperref}
\usepackage{xcolor}
\usepackage{amssymb, amsmath}

\begin{document}


\title{Effective mass approach to memory in non-Markovian systems}

\author{M. Wi\'{s}niewski}
\affiliation{Institute of Physics, University of Silesia, 41-500 Chorz{\'o}w, Poland}
\author{J. {\L}uczka}
\affiliation{Institute of Physics, University of Silesia, 41-500 Chorz{\'o}w, Poland}
%
%
\author{J. Spiechowicz}
\affiliation{Institute of Physics, University of Silesia, 41-500 Chorz{\'o}w, Poland}
\email{jakub.spiechowicz@us.edu.pl}

\begin{abstract}
Recent pioneering experiments on non-Markovian dynamics done e.g. for active matter have demonstrated that our theoretical understanding of this challenging yet hot topic is rather incomplete and there is a wealth of phenomena still awaiting discovery. It is related to the fact that typically for simplification the Markovian approximation is employed and as a consequence the memory is neglected. Therefore methods allowing to study memory effects are extremely valuable.
We demonstrate that a non-Markovian system described by the Generalized Langevin Equation (GLE) for a Brownian particle of mass $M$ can be approximated by the memoryless Langevin equation in which the memory effects are correctly reproduced solely via the effective mass $M^*$ of the Brownian particle which is determined only by the form of the memory kernel. Our work lays the foundation for an impactful approach which allows to readily study memory-related corrections to Markovian dynamics. 
\end{abstract}

\maketitle

\section{Introduction}
The role of memory in dynamics of systems is an issue which seems to attract everlasting activity in multiple contexts reaching even far beyond the scope of physics. Recent renaissance of this conundrum is attributed to active matter \cite{kanazawa2020, banerjee2022, militaru2021, narinder2018, tucci2022, cao2023}, spin glasses \cite{jesi-baity2023}, protein-folding kinetics \cite{netz}, random walk theory \cite{levenier2022,alessandro2021,guerin2016,barbier2022}, search strategies \cite{meyer2021,levenier2020}, animal mobility \cite{vilk2022}, nonlinear fluctuation-dissipation relation \cite{engbring2023}, quantum stochastic processes \cite{milz2020,milz2021}, quantum simulations \cite{white2020, wu2022} and tomography \cite{white2022}, to name only a few. Understanding the role of memory in physics therefore appears both as a hot topic and a major challenge.

The origin of memory in system dynamics is typically related to its complexity, in particular due to a large number of degrees of freedom. It may be an effect of properties of the system itself such as viscoelasticity \cite{goychuk2022,ginot2022,ferrer2020,gomez2016,goychuk2012} or emerge as a result of interplay with environment such as hydrodynamic interactions between the system and the surrounding fluid in its immediate vicinity \cite{franosch2011, huang2010, goychuk2019}. It can result also from either external or internal nonequilibrium noise \cite{martineau2022,hanggi1995,luczka2005}.

The emanation of memory is typically assisted by the emergence of correlated thermal fluctuations. However, to simplify the system one often employs the Markovian approximation in which they are modeled  as the $\delta$-correlated Gaussian process and consequently its dynamics is described by the memoryless equation. However, this idealization has little in common with physical reality. In particular, even at the deep fundamental level of quantum realm energy of the system coupled to thermal vacuum diverges when the Markovian approximation is imposed \cite{spiechowicz2021scirep}. Consequently, ''\emph{non-Markov is the rule, Markov is the exception}'' \cite{vankampen}.

Markovian dynamics is in principle completely characterized if the transition probability distribution and the initial state of the system is known. In contrast, non-Markovian dynamics is completely described by an infinite set of multidimensional probability distributions which cannot be determined from the lower dimensional ones. This fact reveals that analysis of non-Markovian systems is much more difficult to handle even for selected cases. Moreover, it explains why recent pivotal results on dynamics with presence of memory have been often first discovered with pioneering experiments and only later explained theoretically. Therefore methods allowing to investigate memory effects are extremely valuable since our understanding of the non-Markovian dynamics is rather incomplete and there is a wealth of phenomena still awaiting discovery. 

To address this urgent problem, we consider the GLE formalism as a universal framework for investigating the non-Markovian dynamics. For a Brownian particle of mass $M$ 
subjected to a potential $U(x, t)$ and driven by thermal equilibrium fluctuations $\eta(t)$ modeled as a zero-mean stationary Gaussian process the GLE reads \cite{luczka2005}
\begin{equation} \label{eq:gle}
	M\dot{v}(t) + \Gamma \int_0^t K(t-s) v(s) \mathrm{d}s = -U'(x(t), t) + \eta(t),
\end{equation}
where $x(t)$ is a position of the particle at time $t$, \mbox{$v(t)=\dot{x}(t)$} is its velocity and $\Gamma$ stands for the dissipation constant (the friction coefficient). 
The correlation function of $\eta(t)$ is related to the memory kernel $K(t)$ characterized by the memory time $\tau_c$ via the fluctuation-dissipation theorem \cite{kubo1966}, 
\begin{equation} \label{eq:corr_eta}
	\langle \eta(t)\eta(s) \rangle = \Gamma k_B T K(|t-s|). 
\end{equation}
where $k_B$ is the Boltzmann constant and $T$ is temperature of the system. Due to this relation the memory time $\tau_c$ is equivalent to the correlation time of thermal fluctuations. We note that (\ref{eq:gle}) assumes the bilinear coupling between the system and thermostat and is no longer valid if a nonlinear interaction takes place \cite{hadrien}.

Despite several decades of studies in the GLE framework its memoryless version, which nevertheless captures the memory-induced properties of the system, has not been proposed. Our paper is therefore first of its kind to demonstrate that the effects of short memory in non-Markovian dynamics (\ref{eq:gle}) can be absorbed \emph{solely} into \emph{effective mass} of the particle in the corresponding memoryless equation for which the joint process $\{x(t), v(t)\}$ is Markovian. Moreover, our approximation is applicable for a wide class of integrable memory kernels $K(t)$ \cite{davies}, including e.g. both the power-law (with the exponent larger than 2) and the Gaussian decay.

In contrast, another approach to the GLE is to use the Markovian embedding \cite{straub1986,siegle2010,siegle2010pre,klippenstein2021} of the dynamics into the multidimensional Markovian process, in which the evolution of the pair $\{x(t), v(t)\}$ is, however, still non-Markovian. Moreover, this procedure is exact only for a few selected cases while for the others it is a non-unique approximation. If e.g. $K(t) \sim (1 + t/\tau_c)^{-\alpha}$ one has to consider the infinite-dimensional Markovian process \cite{abate} whose analysis is as complicated as the starting non-Markovian one. It can be handled only by arbitrary truncation of the problem dimensionality and therefore the question on the impact of finite dimension effects on the so obtained results always arises. On the other hand, if e.g. $K(t) \sim \exp{[-(t/\tau_c)^2]}$ then the Markovian embedding is impossible.

We note that the idea behind our scheme is similar to the effective mass approach \cite{pekar,kittel} frequently encountered in condensed matter physics. There it describes the mass that the particle (e.g. electron) seems to have under influence of external fields or interactions with other entities. It often helps to radically simplify a complicated system by modeling it as the free particle with the effective mass. Here, the latter allows to replace complex non-Markovian dynamics with much more straightforward Markovian one which, however, still takes into account the memory effects.

The paper is organized as follows. In the next section we derive the effective mass approach for a system described by the GLE. In Sec. III we validate this new method for a fundamental problem of nonequilibrium statistical physics, namely, transport of a driven Brownian particle in a periodic potential and discuss its limitations. Finally, Sec. IV provides a summary and conclusions.

\section{Effective mass approach}

The standard Markovian approximation to the GLE is obtained for the case
\begin{equation}
	K(t) = 2 \delta(t).
\end{equation}
Then Eq. (\ref{eq:gle}) is reduced to the memoryless form
\begin{equation} \label{le}
	M \dot{v}(t) + \Gamma v(t) = -U'(x,t) + \xi(t), 
\end{equation}
where zero-mean thermal noise $\xi(t)$ is a $\delta$-correlated Gaussian process (white noise), 
\begin{equation} \label{eq:corr_xi}
	\langle \xi(t)\xi(s) \rangle = 2\Gamma k_B T \delta(t-s).	
\end{equation}
It means that the memory effects are completely neglected. We want to propose a more refined method for the situation when the memory time (or the correlation time of thermal fluctuations) is short but non-zero. The similar case appears e.g. in investigation of the overdamped dynamics when formally the dimensionless mass of the particle is zero and the strong damping regime for which it is small but non-zero, see Ref. \cite{slapik2018}. 

In the following we consider a class of integrable memory kernels $K(t)$ for which 
\begin{equation} \label{eq:norm_K}
	\int_0^\infty K(t)\mathrm{d}t = 1, \quad  \int_0^\infty t K(t)\mathrm{d}t \ \  \mbox{is finite}. 
\end{equation}
The first integral is related to the finite dissipation (damping) strength, see Eq. (4.17) in Ref. \cite{Ingold}, whereas the second one refers to the finite memory time, see Eq. (4.18) in Ref. \cite{Ingold}.  By virtue of the relation (\ref{eq:corr_eta}) thermal noise correlation function should decay sufficiently fast in the long time limit $t \to \infty$, at least as fast as $K(t) \sim 1/t^{2+\epsilon}$ for a certain $\epsilon > 0$. Examples of such cases are the exponential (the Drude model), Gaussian and algebraic decay with the exponent larger than two \cite{bialasPRA}, to name only a few. Let us mention that similar integrability conditions for the thermostat correlation functions are imposed in the mathematical theory of the weak coupling limit for quantum open systems \cite{davies}. 

We redefine the function $K(t)$ in order to note explicitly its dependence on the memory time $\tau_c$, namely, 
\begin{equation}
	K(t) = \frac{1}{\tau_c} K^*(t/\tau_c). 
\end{equation}
We observe that if $K(t)$ is normalizable to unity then it is so also for $K^*(t)$. Moreover,   $K(t)$ is the Dirac delta sequence on the interval $(0, \infty)$  and in the limit $\tau_c \to 0$ Eq. (\ref{eq:gle}) reduces to Eq. (\ref{le}). 

The integral term in the GLE (\ref{eq:gle}) can be rewritten as 
\begin{align} \label{int}
	I &=  \int_0^t K(s) v(t-s) \mathrm{d}s \nonumber \\ &= \frac{1}{\tau_c} \int_0^{t} K^*(s/\tau_c) v(t-s) \mathrm{d}s \nonumber \\ &= \int_0^{t/\tau_c} K^*(u) v(t- \tau_c u) \mathrm{d}u.
\end{align}
For short but non-zero memory time $\tau_c \neq 0$, when the memory kernel $K(t)$ decays rapidly, we expand the function $v(t-\tau_c u)$ into a Taylor series
\begin{equation}
	v(t-\tau_c u) \approx v(t) - \tau_c u\dot{v}(t)
\end{equation}
and neglect the terms of the order higher than $\tau_c$.
Consequently the leading term in Eq. (\ref{int}) reads
\begin{equation}
	I \approx  v(t)  - \varepsilon \tau_c \dot{v}(t), 
\end{equation}
where the dimensionless parameter $\varepsilon$ is given by the relation
\begin{equation} \label{epsilon}
	\varepsilon  = \int_0^{\infty} u K^*(u)\mathrm{d}u
\end{equation}
for which we extended the upper limit of integration to infinity provided that $\tau_c$ is sufficiently small, i.e. $\int_0^{t/\tau_c} F(u) \mathrm{d}u \approx \int_0^{\infty} F(u) \mathrm{d}u$ for any function $F(u)$. Finally, the original GLE (\ref{eq:gle}) is approximated by the Langevin equation
\begin{equation}
	(M - \varepsilon \tau_c \Gamma  )\dot{v}(t) + \Gamma v(t) = -U'(x,t) + \eta(t)
\end{equation}
Here, the dissipative term proportional to $v(t)$ reads $\Gamma v(t)$ for which the corresponding memory kernel is expressed by the Dirac delta function. By virtue of the fluctuation-dissipation relation (\ref{eq:corr_eta}) one finds that thermal noise $\eta(t)$ in such a case is $\delta$-correlated and the original GLE is approximated by the following equation
\begin{equation} \label{ema}
	M^*\dot{v}(t) + \Gamma v(t) = -U'(x, t) + \xi(t),
\end{equation} 
where $\xi(t)$ is white thermal noise obeying the fluctuation-dissipation relation given by Eq. (\ref{eq:corr_xi}).

The effective mass of the particle is identified as
\begin{equation}
\label{effectM}
	M^* = M - \varepsilon \tau_c \Gamma = M \left(1 - \varepsilon \frac{\tau_c}{\tau_L}\right).
\end{equation}
Eq. (\ref{ema}) means that the non-Markovian dynamics with short memory time can be approximated by the much simpler corresponding Markovian one but with the effective mass of the system. We note that the latter depends on the ratio of two characteristic times $\tau_c/\tau_L$, where $\tau_L = M/\Gamma$ stands for the well-known velocity relaxation time of the free Brownian particle. Moreover, this ratio needs to satisfy the relation $\tau_c/\tau_L < 1/\varepsilon$ so that the effective mass is positive $M^* > 0$. Otherwise the system is non-dissipative and various unphysical effects such as an increase of the particle energy to infinity as time grows may emerge. In particular, for the strict white noise limit $\tau_c \to 0$ the effective mass is equal to the actual mass $M^* = M$, whereas e.g. for an exponentially decaying kernel 
\begin{equation}
	K(t) = \frac{1}{\tau_c}e^{-t/\tau_c}
\end{equation} 
the parameter $\varepsilon = 1$ and the effective mass is
\begin{equation}
	M^* = M \left(1 - \frac{\tau_c}{\tau_L}\right),
\end{equation}
while for the Gaussian decay
\begin{equation}
	K(t) = \frac{2}{\tau_c \sqrt{\pi} }e^{-(t/\tau_c)^2},
\end{equation}
the parameter $\varepsilon = 1/\sqrt{\pi}$ and the effective mass is larger,
\begin{equation}
	M^* =  M \left(1 - \frac{1}{\sqrt{\pi}}\frac{\tau_c}{\tau_L}\right).
\end{equation}

The effective mass approach can be derived also in the Laplace space to obtain its interesting interpretation. For this purpose we focus on the left hand side $L$ of the original GLE (\ref{eq:gle}). Its Laplace transform reads
\begin{align} \label{laplace}
	\mathcal{L}\{L\}(z) &= \mathcal{L}\left\{M\dot{v}(t) + \Gamma \int_0^t K(t-s) v(s) \mathrm{d}s\right\}(z) \nonumber \\ &= M z \mathcal{L}\{\dot{v}\}(z) + \Gamma \mathcal{L}\{K\}(z) \mathcal{L}\{v\}(z),
\end{align}
where the transform itself is defined for any function $F(t)$ and the complex number $z$ as
\begin{equation}
	\mathcal{L}\{F\}(z) = \int_0^\infty F(t) e^{-zt} dt.
\end{equation}
The Laplace transform of the memory kernel reads
\begin{align}
	\mathcal{L}\{K\}(z) &= \int_0^{\infty} \frac{1}{\tau_c} K^*(t/\tau_c) \mbox{e}^{-zt} dt \nonumber \\ &= \int_0^{\infty} K^*(u) \mbox{e}^{-\tau_c zu} du \nonumber \\ &= \int_0^{\infty} K^*(u) [1-\tau_c z u + ...] du \nonumber \\ &\approx  1 - \varepsilon \tau_c z, 
\end{align}
where $\varepsilon$ is defined in Eq. (\ref{epsilon}) and as previously we neglected terms of the order higher than $\tau_c$ provided that it is small. Inserting this formula into Eq. (\ref{laplace}) gives
\begin{equation}
	\mathcal{L}\{L\}(z) = (M - \varepsilon \tau_c \Gamma) z \mathcal{L}\{\dot{v}\}(z) + \Gamma \mathcal{L}\{v\}(z).
\end{equation}
The inverse Laplace transform corresponds to the left hand side of Eq. (\ref{ema}). This method gives another interpretation of the effective mass. Since the memory kernel $K(t)$ characterizes the two-point correlation function of thermal fluctuations its transform $\mathcal{L}\{K\}(z)$ is related to their power spectrum. We therefore infer that the leading memory correction to the power spectrum can be interpreted as a correction to the particle mass.
\begin{figure}[t]
    \centering
    \includegraphics{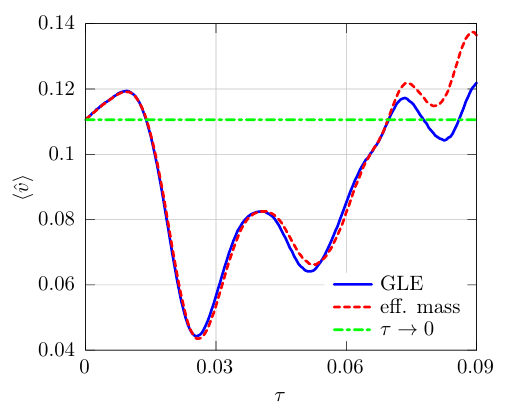}
    \caption{The average velocity $\langle \hat{v} \rangle$ of a Brownian particle as a function of the correlation time $\tau$ for $f=0.1$ obtained from (i) the GLE (\ref{eq:gle_dimless}), (ii) the effective mass approach (\ref{ema_dimless}) and (iii) the standard Markovian approximation $\tau \to 0$ when $m^* = m$. The parameters are $\{m = 1$, $a = 15$, $\omega = 4$, $f = 0.1$, $D = 10^{-3}\}$.}
    \label{fig:v_t}
\end{figure}

\section{Validation and limitation}
It is often argued that if $\tau_c$ is much smaller than other characteristic time scales in the system the impact of memory is negligible and the Markovian approximation $\tau_c \to 0$ can be applied. It allows to radically simplify the underlying analysis. However, we now demonstrate that even when $\tau_c \neq 0$ is the smallest time scale of the system, naive use of the Markovian approximation $\tau_c \to 0$ can give completely wrong results as the influence of short memory may be still prominent. 
In contrast, we show that our scheme offers the same advantages as the Markovian approximation but the impact of short memory can be absorbed into the effective mass $M^*$ of the system and consequently it yields correct predictions. In order to validate our approach we compare three results obtained from: (i)  Eq. (\ref{eq:gle}) for small $\tau_c > 0$, (ii) the effective mass method (\ref{ema}) with $M^* < M$ and (iii) the standard Markovian approximation $\tau_c \to 0$ when $M^* = M$.

We limit ourselves to the situation in which the standard method of the GLE handling in the form of the Markovian embedding is exact, so that the results obtained in this way can serve as a reference for validation of the effective mass approach. The simplest case meeting this requirement is the exponentially decaying memory kernel $K(t) = 1/\tau_c \exp{(-t/\tau_c)}$ for which the Markovian embedding allows to convert the original GLE (\ref{eq:gle}) into a set of ordinary stochastic differential equations \cite{luczka2005}. Let us define the auxiliary stochastic process $w(t)$ via the relation
\begin{equation}
w(t) = \frac{\Gamma}{\tau_c} \int_0^t \mbox{e}^{-(t-s)/\tau_c} v(s)\;ds.
\end{equation}
Then Eq. (\ref{eq:gle}) is transformed into the equivalent form
\begin{subequations} \label{embedd}
\begin{align} 
M \dot v(t) &= - U'(x(t), t) -w(t) + \eta (t), \\
\dot x(t) &=v(t), \\
\dot w(t) &= -\frac{1}{\tau_c} w(t) +\frac{\Gamma}{\tau_c} v(t), \\
\dot \eta (t) &= -\frac{1}{\tau_c} \eta (t) + \frac{1}{\tau_c} \xi(t),
\end{align}
\end{subequations}
where the zero-mean Gaussian white noise $\xi(t)$ obeys $\langle\xi(t)\xi(s)\rangle=2\Gamma k_B T\delta(t-s)$  and the last equation of this set describes the Ornstein-Uhlenbeck noise with the exponential correlation function. 

As a system of interest we pick a driven Brownian particle in a periodic potential. For such a case the potential reads
\begin{equation} \label{potential}
	U(x, t) = V_0  \sin(2\pi x/L) - \left[ A\cos{(\Omega t)} + F \right]x, 
\end{equation}
where $V_0$ is half of the barrier height of the periodic potential with the spatial period $L$. $A\cos{(\Omega t})$ represents the external driving of amplitude $A$ and angular frequency $\Omega$ while $F$ is a static bias which breaks the spatial symmetry of the system and induces the directed transport. This model constitutes a fundamental problem of nonequlibrium statistical physics appearing in numerous contexts including normal and anomalous transport \cite{ros2005, machura2007, nagel2008, spiechowicz2014pre, slapik2019, spiechowicz2019njp, wisniewski2022, wisniewski2023}. The quantity of interest for a present study will be the average velocity of the Brownian particle reading
\begin{equation}
    \langle v \rangle = \lim\limits_{t\to\infty}\frac{1}{t}\int_0^t \langle \dot{x}(s) \rangle \mathrm{d}s,
\end{equation}
where the brackets $\langle \cdot \rangle$ indicate the average over the initial conditions and realizations of thermal noise $\eta(t)$. The former is mandatory for the deterministic dynamics when ergodicity of the system may be broken and the results can be affected by the specific choice of initial conditions \cite{spiechowicz2016scirep,spiechowicz2022entropy}.

The set of ordinary stochastic differential equations (\ref{embedd}) with a non-linear, time-dependent potential given by Eq. (\ref{potential}) cannot be solved analytically. For this reason we had to resort to numerical computations using CUDA environment on modern desktop graphics processing unit. This approach allowed us to accelerate calculations by several orders of magnitude as compared to standard methods \cite{spiechowicz2015cpc}. We employed a weak second-order predictor corrector algorithm to simulate the corresponding dynamics. 
\begin{figure}[t]
    \centering
    \includegraphics{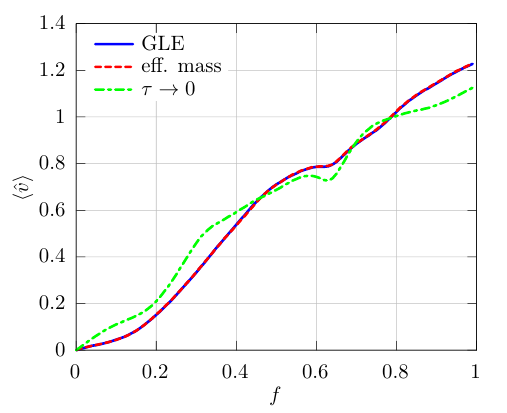}[t]
    \caption{The average velocity $\langle \hat{v} \rangle$ of the Brownian particle as a function of the static bias $f$ for $\tau=0.025$. Other parameters are the same as in Fig. 1.}
    \label{fig:v_f}
\end{figure}

In doing so we first transformed the GLE (\ref{eq:gle}) into the dimensionless form. We introduce the dimensionless position and time
\begin{equation}
	\hat{x} = \frac{x}{L},\quad \hat{t} = \frac{t}{\tau_0}, \quad \tau_0 = \frac{\Gamma L^2}{V_0}.
\end{equation}
The rescaled potential takes the form
\begin{equation}  \label{potential_dimless}
	\hat{U}(\hat{x}, \hat{t}) =  \hat{V}_0(\hat{x}) - \hat{x}\left[ a\cos(\omega\hat{t}) + f \right],
\end{equation}
where
\begin{equation}
	\hat{V}_0(\hat{x}) = \sin(2\pi\hat{x}), \; a = \frac{L}{V_0} A, \; \omega = \tau_0 \Omega, \; f = \frac{L}{V_0} F.
\end{equation}
Thermal noise transforms according to
\begin{equation}
	\hat{\eta}(\hat{t}) = \frac{L}{V_0}\eta(\tau_0 \hat{t}).
\end{equation}
The exponentially decaying memory kernel scales as
\begin{equation}
	\hat{K}(|\hat{t}-\hat{s}|) = \frac{1}{\tau}e^{-|\hat{t}-\hat{s}|/\tau}, \quad \tau = \frac{\tau_c}{\tau_0}, 
\end{equation}
where $\tau$ is the dimensionless correlation time of thermal fluctuations.
Finally, the rescaled mass reads
\begin{equation}
	m = \frac{M}{\tau_0 \Gamma}.
\end{equation}
With such a choice of length and time scales, the dimensionless friction coefficient $\gamma = 1$.

The original GLE (\ref{eq:gle}) describing the non-Markovian dynamics after such a scaling procedure is transformed to the form
\begin{equation} \label{eq:gle_dimless}
	m\dot{\hat v}({\hat t}) + \int_0^{\hat t}{\hat K}({\hat t} -{\hat s}) {\hat v}({\hat s}) \mathrm{d}{\hat s} = -{\hat U}'({\hat x}, {\hat t}) + {\hat \eta}({\hat t}).
\end{equation}
The fluctuation-dissipation relation now is
\begin{equation}
	\langle {\hat \eta} ({\hat t}) {\hat \eta}({\hat s}) \rangle = D {\hat K}(|{\hat t}-{\hat s}|), \quad D = \frac{k_B T}{V_0}. 
\end{equation}
The corresponding effective mass approach reads 
\begin{equation} \label{ema_dimless}
    m^*\dot{{\hat v}}({\hat t}) + {\hat v}({\hat t}) = -{\hat U}'({\hat x}, {\hat t}) + {\hat \xi}({\hat t}), 
\end{equation}
where the rescaled effective mass
\begin{equation}
m^* = m - \tau = \frac{1}{\tau_0}(\tau_L - \tau_c)
\end{equation}
is a difference of the dimensionless mass $m$ and the memory time $\tau = \tau_c/\tau_0$. It can be represented also as a difference of two characteristic time scales $\tau_L$ and $\tau_c$ in units of the third characteristic time $\tau_0 = \Gamma L^2/V_0$ describing the interval in which the overdamped particle moves from the maximum to minimum of the spatially periodic part of the potential (\ref{potential_dimless}). Last but not least, thermal noise is $\delta$-correlated, i.e.
\begin{equation}
	\langle {\hat \xi}({\hat t}) {\hat \xi}({\hat s}) \rangle = 2D \delta({\hat t}-{\hat s}). 
\end{equation}
The dimensionless average velocity of the Brownian particle $\langle \hat{v} \rangle = (\tau_0/L) \langle v \rangle$ was averaged over the ensemble of $2^{16} = 65536$ thermal noise realizations as well as the initial conditions $\hat{x}(0)$ and $\hat{v}(0)$ distributed uniformly over the interval $[0,1]$ and $[-2;2]$, respectively.  Each trajectory of the system was simulated up to the final time $t_f = 2 \times 10^4 \times \hat{\tau}_{\omega}$ with a time step $h = 10^{-3} \times \hat{\tau}_{\omega}$ where $\hat{\tau}_{\omega} = 2\pi/\omega$ stands for the period of the external driving force $a \cos{(\omega \hat{t})}$.
\begin{figure}[t]
    \centering
    \includegraphics{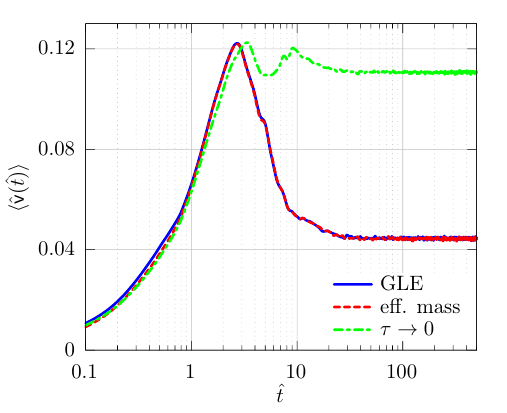}
    \caption{Time evolution of the mean velocity averaged over the external driving period $\langle \hat{\mathsf{v}} \rangle$ for $\tau=0.025$ and $f=0.1$. Other parameters are the same as in Fig. 1}
    \label{fig:transient}
\end{figure}

In the following part of the paper we consider the exemplary parameter regime: $\{m = 1$, $a = 15$, $\omega = 4$, $f = 0.1$, $D = 10^{-3}\}$. 

In Fig.~\ref{fig:v_t} we show the average velocity $\langle \hat{v} \rangle$ of the Brownian particle as a function of the memory time $\tau$ for the static bias $f = 0.1$. We compare three results obtained from (i) the dimensionless GLE (\ref{eq:gle_dimless}) via the exact Markovian embedding, (ii) the effective mass approach with $m^* < m$ (\ref{ema_dimless}) and (iii) the standard Markovian approximation $\tau \to 0$ when $m^* = m$. The first observation is that although the memory time $\tau$ is two orders of magnitude smaller than other characteristic times of the system like $\hat{\tau}_L = m = 1$ or $\tau_0 = 1$ the standard Markovian approximation completely fails to predict the average velocity of the particle $\langle \hat{v} \rangle$. One often claims that in such a case this simplification can be done. This example shows that it is not true in general and a special caution is needed even when the memory time $\tau$ is much smaller than the other time scales. In contrast, the average velocity $\langle \hat{v} \rangle$ calculated using the effective mass approach perfectly follows the solution obtained from the full GLE up to $\tau \approx 0.07$. The correctness of the effective mass approach is stable over the variation of the system parameters. In Fig.~\ref{fig:v_f} we present the average velocity $\langle \hat{v} \rangle$ as a function of the static bias $f$ for the memory time $\tau = 0.025$. Again, the studied system in the Markovian limit $\tau \to 0$ fails to correctly predict the behavior of the non-Markovian system, however, the characteristic obtained for the effective mass approach perfectly fits the original curve. 

Moreover, this equivalence is not limited only to the asymptotic long time state of the system but it is preserved also in the transient regime. Let us define the velocity averaged over the period $\mathsf{T} = 2\pi/\omega$ of the external driving force 
\begin{equation}
    \hat{\mathsf{v}}(\hat{t}) = \frac{1}{\mathsf{T}} \int_{\hat{t}}^{\hat{t}+\mathsf{T}} \hat{v}(s)\mathrm{d}s.
\end{equation}

In Fig.~\ref{fig:transient} we show how the velocity $\langle \hat{\mathsf{v}}(\hat{t}) \rangle$ evolves from the initial state of system to its asymptotic regime. The curve corresponding to the effective mass approach agrees with the response of the non-Markovian system modeled by the GLE. This fact must be contrasted with the standard Markovian approximation $\tau \to 0$ which is not correct in both the intermediate and asymptotic situation.

\begin{figure}[t]
    \centering
    \includegraphics{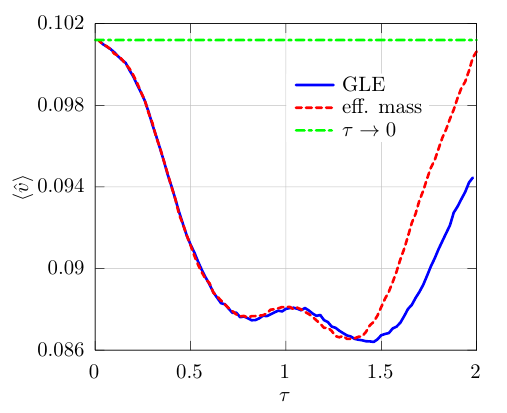}
    \caption{The average velocity $\langle \hat{v} \rangle$ of the Brownian particle as a function of the memory time $\tau$ for $m=149$, $a=50$, $\omega=0.2$, $f=0.1$ and $D=0.01$. The effective mass approach is correct up to $\tau \approx \tau_0 = 1$.}
    \label{fig:v_long_tau}
\end{figure}
Finally, we would like to note that the range of applicability of the effective mass approach is not restricted to the case of short memory time $\tau \ll 1$. What matters is the relation of the latter to the characteristic time describing the velocity relaxation $\hat{\tau}_L = m$. When it is long, i.e. $m \gg 1$, then the effective mass approach can be correct even for the memory time of the order of the other time scales, e.g. $\tau \approx \tau_0 = 1$. In Fig.~\ref{fig:v_long_tau} we illustrate such a case. The standard Markovian approximation is defined for $\tau \to 0$ when the memory time is much smaller than the other time scales of the system. Therefore one would not expect that the non-Markovian system could be approximated by the Markovian model when it is not the case. In contrast, we show that the mass correction in the effective mass approach correctly reproduces the memory effects in the Markovian model even when the correlation time $\tau$ is of the order of the other characteristic time scales such as $\tau_0$.

\section{Summary}
In conclusion, we presented a novel approximation that transforms the non-Markovian dynamics in presence of short memory into Markovian one which captures the memory-induced properties of the system. Effects of short memory are reflected there solely in effective mass of the particle which is determined only by the form of the memory kernel or, equivalently, by the correlation function of thermal fluctuations. It implies that complexity of the underlying dynamics can be radically reduced by (i) exploiting the correspondence between the memory and mass correction to significantly limit the parameter space of the problem and (ii) transforming the stochastic integro-differential into the stochastic differential equation.

This approach works universally for a wide class of integrable memory kernels provided that the memory time is much shorter than the characteristic time scale describing the velocity relaxation. Therefore our work lays the foundation for impactful methodology which allows to study corrections to Markovian dynamics resulting from correlations or memory in a vast number of systems described by the Generalized Langevin Equation. These often neglected memory corrections can radically change the system behavior. Since the effective mass approach makes  investigation of the role of memory much easier and accessible we expect the emergence of vibrant follow up works with novel insights on non-Markovian dynamics as well as new memory-induced effects.

\section*{Acknowledgment}
This work has been supported by NCN Grant No. 2022/45/B/ST3/02619 (J.S.).


\begin{thebibliography}{99}
	\bibitem{kanazawa2020} K. Kanazawa, T. G. Sano, A. Cairoli and A. Baule, Nature 579, 364 (2020)
	\bibitem{banerjee2022} J. P. Banerjee, R. Mandal, D. S. Banerjee et al., Nat. Commun. 13, 4533 (2022)
	\bibitem{militaru2021} A. Militaru, M. Innerbichler, M. Frimmer, F. Tebbenjohanns, L. Novotny and C. Dellago, Nat. Commun. 12, 2446 (2021)
	\bibitem{narinder2018} N. Narinder, C. Bechinger and J. R. Gomez-Solano, Phys. Rev. Lett. 121, 078003 (2018)
	\bibitem{tucci2022} G. Tucci, E. Roldan, A. Gambassi, R. Belousov, F. Berger, R. G. Alonso and A. J. Hudspeth, Phys. Rev. Lett. 129, 030603 (2022)
	\bibitem{cao2023} X. Cao, D. Das, N. Windbacher et al. Nat. Phys. (2023) https://doi.org/10.1038/s41567-023-02213-1
	
	\bibitem{jesi-baity2023} M. Jesi-Baity, E. Calore, A. Cruz et al. Nat. Phys. 19, 978 (2023)
	\bibitem{netz} C. Ayaz, L. Tepper, F. N. Brünig, J. Kappler, J. O. Daldrop and R. R. Netz, 
PNAS 118,  e2023856118 (2021) 

	\bibitem{levenier2022} N. Levenier, T. V. Mendes, O. Benichou, R. Voituriez and T. Guerin, Nat. Commun. 13, 5319 (2022)
	\bibitem{alessandro2021} J. d'Allesandro, A. Barbier-Chebbah, V. Cellerin, O. Benichou, R. M. Mege, R. Voituriez and B. Ladoux, Nat. Commun. 12, 4118 (2021)
	\bibitem{guerin2016} T. Guerin, N. Levenier, O. Benichou and R. Voituriez, Nature 534, 356 (2016)
	\bibitem{barbier2022} A. Barbier-Chebbah, O. Benichou and R. Voituriez, Phys. Rev. X 12, 011052 (2022)
	
	\bibitem{meyer2021} H. Meyer and H. Rieger, Phys. Rev. Lett. 127, 070601 (2021)	
	\bibitem{levenier2020} N. Levenier, J. Textor, O. Benichou and R. Voituriez, Phys. Rev. Lett. 124, 080601 (2020)
	
	\bibitem{vilk2022} O. Vilk, D. Campos, V. Mendez, E. Lourie, R. Nathan and M. Assaf, Phys. Rev. Lett. 128, 148301 (2022)
	\bibitem{engbring2023} K. Engbring, D. Boriskovsky, Y. Roichmann and B. Lindner, Phys. Rev. X 13, 021034 (2023)
	\bibitem{milz2020} S. Milz, D. Egloff, P. Taranto, T. Theurer, M. B. Plenio, A. Smirne and S. F. Huelga, Phys. Rev. X 10, 041049 (2020)
	\bibitem{milz2021} S. Milz and K. Modi, PRX Quantum 2, 030201 (2021)
	\bibitem{white2020} G. A. L. White, C. D. Hill, F. A. Pollock, L. C. L. Hollenberg and K. Modi, Nat. Commun. 11, 6301 (2020)
	\bibitem{wu2022} K. D. Wu, C. Yang, R. D. He, Nat. Commun. 14, 2624 (2023)
	\bibitem{white2022} G. A. L. White, F. A. Pollock, L. C. L. Hollenberg, K. Modi and C. D. Hill, PRX Quantum 3, 020344 (2022)
	\bibitem{goychuk2022} I. Goychuk, PNAS 119, e2205637119 (2022)
	\bibitem{ginot2022} F. Ginot, J. Caspers, M. Kruger and C. Bechinger, Phys. Rev. Lett. 128, 028001 (2022)
	\bibitem{ferrer2020} B. R. Ferrer, J. R. Gomez-Solano and A. V. Arzola, Phys. Rev. Lett. 126, 108001 (2021)
	\bibitem{gomez2016} J. R. Gomez-Solano, A. Blokhuis and C. Bechinger, Phys. Rev. Lett. 116, 138301 (2016)
	\bibitem{goychuk2012} I. Goychuk, Adv. Chem. Phys. 150, 187 (2012)
	\bibitem{franosch2011} T. Franosch, M. Grimm, M. Belushkin et al., Nature 478, 85 (2011)
	\bibitem{huang2010} R. Huang, I. Chavez, K. Taute et al., Nat. Phys. 7, 576 (2011)
	\bibitem{goychuk2019} I. Goychuk, Phys. Rev. Lett. 123, 180603 (2019) 
	\bibitem{martineau2022} S. Martineau, T. Saffold, T. T. Chang and H. Ronellenfitsch, Phys. Rev. Lett. 128, 098301 (2022)
	
	\bibitem{hanggi1995} P. Hanggi and P. Jung, Adv. Chem. Phys. 89, 239 (1995)
	\bibitem{luczka2005} J. {\L}uczka, Chaos 15, 026107 (2005)
	\bibitem{spiechowicz2021scirep} J. Spiechowicz and J. {\L}uczka, Sci. Rep. 11, 4088 (2021)
	\bibitem{vankampen} N. G. van Kampen, Braz. J. Phys. 28, 90 (1998)
	
	\bibitem{kubo1966} R. Kubo, Rep. Prog. Phys. 29, 255 (1966)
	\bibitem{hadrien} H. Vroylandt, P. Monmarche, J. Chem. Phys. 156, 244105 (2022)

	
	\bibitem{davies} E. B. Davies, Commun. Math. Phys. 39, 91 (1974)
	\bibitem{straub1986} J. E. Straub, M. Borkovec, and B. J. Berne, J. Chem. Phys. 84, 1788 (1986)
	\bibitem{siegle2010} P. Siegle, I. Goychuk and P. H\"anggi, Phys. Rev. Lett. 105, 100602 (2010)
	\bibitem{siegle2010pre} P. Siegle, I. Goychuk and P. H\"anggi, Phys. Rev. E 81, 011136 (2010)
	\bibitem{klippenstein2021} V. Klippenstein, M. Tripathy, G. Jung, F. Schmid and N. F. A. van der Vegt, J. Phys. Chem. B 125, 4931 (2021)
	\bibitem{abate} J. Abate and W. Whitt. J., J. Oper. Res. Soc. Japan 42, 268 (1999)


	\bibitem{pekar} S. Pekar, Zh. Eksp. Teor. Fiz. 16, 933 (1946)
	\bibitem{kittel} Ch. Kittel, Introduction to Solid State Physics (New York, Wiley, 1996)
	\bibitem{slapik2018} A. S{\l}apik, J. {\L}uczka, and J. Spiechowicz, Commun. Nonlinear Sci. Numer. Simul. 55, 316 (2018)
	\bibitem{Ingold} T. Dittrich, P. H\"anggi, G.-L. Ingold, B.  Kramer, G. Sch\"on, W. Zwerger,  Quantum Transport and Dissipation (Wiley, Weinheim, 1998), see Chap. 4 therein
	\bibitem{bialasPRA} J. Spiechowicz, P. Bialas and J. {\L}uczka, Phys. Rev. A 98, 052107 (2018)
	
	\bibitem{ros2005} A. Ros, R. Eichhorn, J. Regtmeier, T. T. Duong, P. Reimann, and D. Anselmetti,  Nature 436, 928 (2005)
	\bibitem{machura2007} L. Machura, M. Kostur, P. Talkner, J. Luczka, and P. Hanggi, Phys. Rev. Lett. 98, 040601 (2007)
	\bibitem{nagel2008} J. Nagel, D. Speer, T. Gaber, A. Sterck, R. Eichhorn, P. Reimann, K. Ilin, M. Siegel, D. Koelle, and R. Kleiner, Phys. Rev. Lett. 100, 217001 (2008)
	\bibitem{spiechowicz2014pre} J. Spiechowicz, P. Hanggi, and J. Luczka, Phys. Rev. E 90, 032104 (2014)
	\bibitem{slapik2019} A. Słapik, J. Łuczka, P. Hanggi, and J. Spiechowicz, Phys. Rev. Lett. 122, 070602 (2019)
	\bibitem{spiechowicz2019njp} J. Spiechowicz, P. Hänggi, and J. Luczka, New J. Phys. 21, 083029 (2019)
	\bibitem{wisniewski2022} M. Wi\'{s}niewski and J. Spiechowicz, New J. Phys. 24, 063028 (2022)
	\bibitem{wisniewski2023} M. Wi\'{s}niewski and J. Spiechowicz, Chaos 33, 063114 (2023)
	\bibitem{spiechowicz2016scirep} J. Spiechowicz, J. Luczka and P. Hanggi, Sci. Rep. 6, 30948 (2016)
	\bibitem{spiechowicz2022entropy} J. Spiechowicz, P. Hanggi and J. Luczka, Entropy 24, 98 (2022)
	\bibitem{spiechowicz2015cpc} J. Spiechowicz, M. Kostur and L. Machura, Comp. Phys. Commun. 191, 140 (2015)
\end{thebibliography}
\bibliographystyle{unsrt}

\end{document}